\documentstyle[11pt,IAU207_pasp,twoside,psfig]{article}
\def\edcomment#1{\iffalse\marginpar{\raggedright\sl#1\/}\else\relax\fi}
\reversemarginpar

\begin{document}
\title{The RR Lyrae variables in M54 and the Sgr dwarf galaxy}
\author{C. Cacciari, M. Bellazzini}
\affil{Osservatorio Astronomico, Via Ranzani 1, 40127 Bologna, Italy}
\author{S. Colucci}
\affil{Dip. di Astron., Univ. di Bologna, Via Ranzani 1, 40127 Bologna, Italy}

\begin{abstract}
We report on new  B, V and I CCD photometry of the globular 
cluster M54 that was aimed at the study of its variable stars. 
With respect to the previous most recent work on M54 we have nearly doubled 
the number of detected variable stars: M54 can now be classified as 
intermediate in the Oosterhoff groups. The metallicity distribution 
for the cluster and the Sgr dSph field is obtained from the red giant 
stellar population, and for the variables.  
\end{abstract}

\section{Introduction}

Our data consist of 54 B, 57 V and 52 I frames taken at the 1.54cm ESO-Danish 
telescope in July 1999. The data reduction was performed using 
the ISIS package (Alard 2000), which is based on 
the method of image subtraction. This technique is particularly powerful when 
searching for variability in crowded fields, both in terms of detection rate 
and in terms of photometric accuracy. 
As an example, in Fig. 1 we show for comparison the V light curves of 3 RR 
Lyrae variables that were measured by Layden \& Sarajedini 2000 (hereafter
LS2000) with traditional techniques (PSF-fitting) and by us with ISIS. 

\begin{figure}
\centerline{\vbox{
\psfig{figure=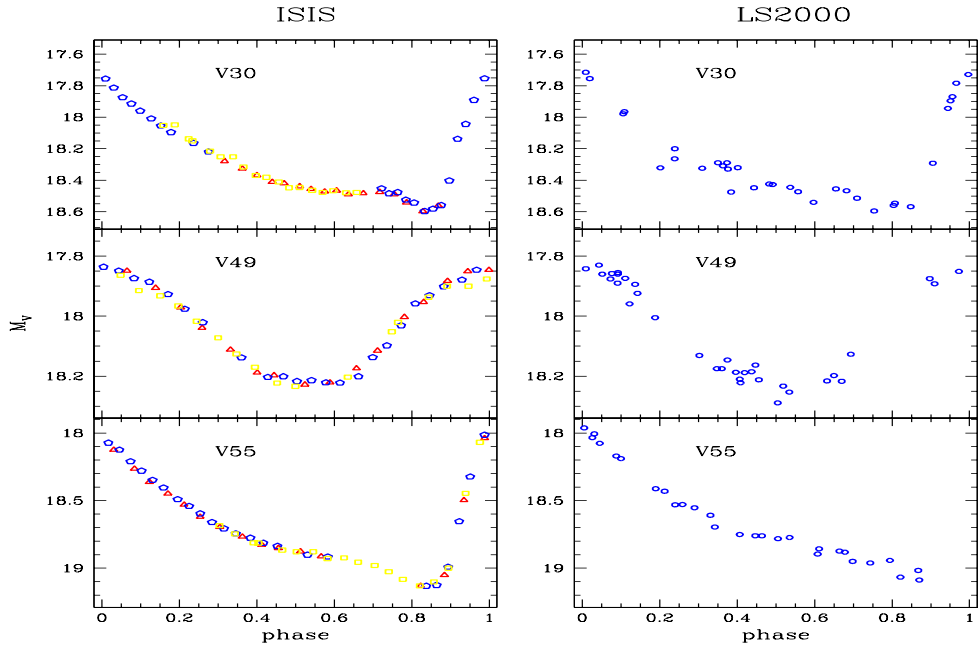,width=14.0cm,height=9.0cm}
}}
\caption{Comparison between our results (ISIS reduction package) and LS2000 
(DoPhot) on three common stars.}
\end{figure} 

\section{The RR Lyrae variables}

With respect to the previous study by LS2000, the number of detected variable 
stars has increased from 117 to 211, in particular more small amplitude 
variables (RRc and long period RRab) have been found. {\em M54 can now be 
classified as intermediate in the Oosterhoff groups}. 

The reddening can be estimated from the (V--I) colors at minimum light of 
the RRab-type pulsators and is E(V--I)=0.17 $\pm$ 0.02 mag. Assuming 
[Fe/H]=--1.55 for M54, and the relation Mv(RR) = 0.20 [Fe/H] + 0.98 (Fernley 
et al. 1998), we obtain a distance modulus (m-M)$_0$ = 17.07 for M54,  
in a distance scale where the LMC has a true distance modulus 18.44. 

The Fourier decomposition of the V light curves provides information on 
several physical parameters (e.g. metallicity, see below). 

\subsection{The metallicity distribution}

We have obtained an improved CMD, where one can clearly identify the 
contributions of the Galactic disk and bulge contaminating components, 
the M54 Red Giant Branch (RGB), and the Sgr intermediate and metal-rich RGBs.
 
For the RGB stellar components the metallicity has been estimated by 
comparison with template RGB ridge lines of Galactic globular clusters at 
various metallicities (Saviane et al. 2000). 
For the RR Lyraes the metallicity has been estimated by Fourier decomposition 
of the V light curves (Kovacs \& Jurcsik 1997).  
In Fig. 2 we show  the metallicity distributions for the RGB stars of 
the Sgr field (about half a degree away from M54 - Bellazzini et al. 1999),  
the RGB stars in the 
field of M54, and  the RR Lyrae variables in the field of M54. 

$\bullet$ The RGB stars in the Sgr field show two clear components peaking at
[Fe/H]  about --0.6 and  --1.55 and the hint of a component at about --2.1. 

$\bullet$ In the field centered on M54 one can identify the same components 
as above, at [Fe/H] = --1.55 and  --0.6, albeit with different relative 
proportions.  The hint of the component at --2.1 is not seen, possibly 
because included in the decontamination from the 
Galactic disk/bulge contribution.  
A small but detectable intermediate Sgr population appears at about --1.2 .  

$\bullet$ The RR Lyrae stars belong to populations at three different 
metallicities, i.e. --1.55 (produced by M54 and the Sgr field), and --1.2 
and --2.1 (produced by the Sgr field only). The metal-rich Sgr field 
population at --0.6,  as expected, does not  produce RR Lyraes.  

The paper with the final results is in preparation.

\begin{figure}
\centerline{\vbox{
\psfig{figure=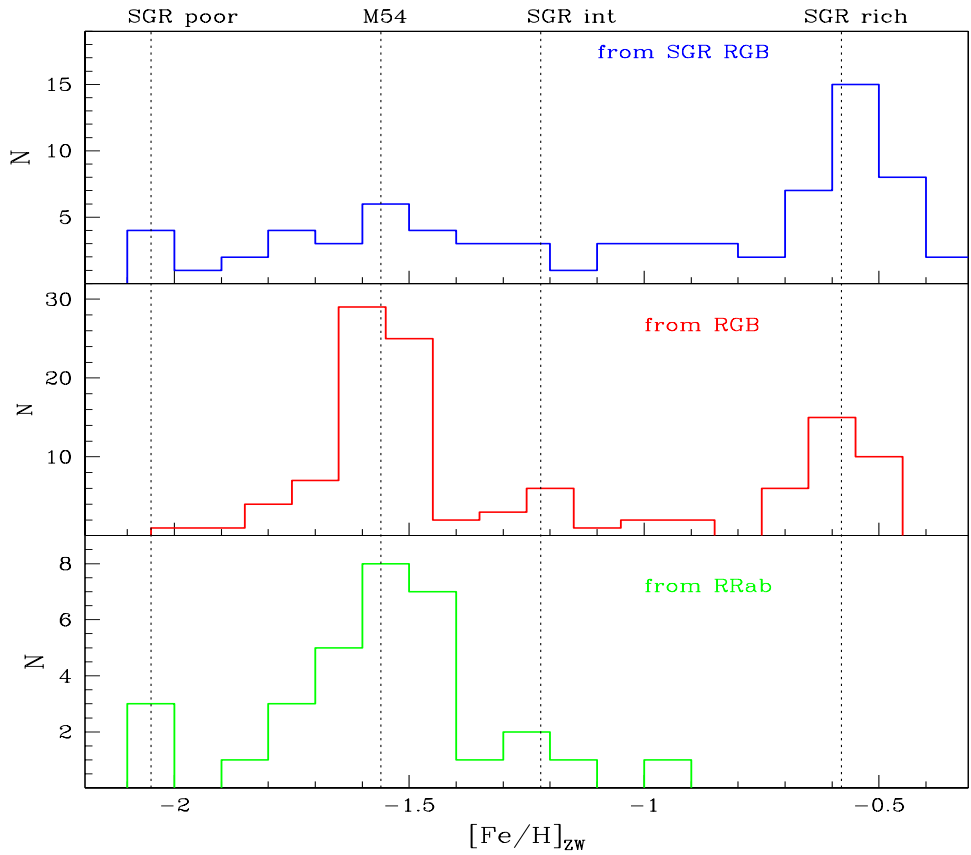,width=14.0cm,height=12.0cm}
}}
\caption{Metallicity distributions. Upper panel: RGB stars in the Sgr field
half a degree away from M54. Middle panel: RGB stars in the M54 field.
Lower panel: RR Lyrae variables in the M54 field.}
\end{figure}

\end{document}